\documentclass[prl,twocolumn,twoside,showpacs,preprintnumbers,amsmath,amssymb,floatfix,aps]{revtex4}

\usepackage{graphicx}% Include figure files
\usepackage{dcolumn}% Align table columns on decimal point
\usepackage{bm}% bold math

\renewcommand{\vec}[1]{\bm{#1}}

\newcommand{\pr}[1]{{\sc{\lowercase{#1}}}}

\newcommand{\codeversion}{1.00}

\newcommand{\vu}{v}
\newcommand{\vau}{t}
\newcommand{\disregard}[1]{}
\newcommand{\rmd}{{\rm d}}

\newcommand{\short}[1]{}

\begin{document}

\title{Convergence of density-matrix expansions for nuclear interactions}

\author{ B.G.~Carlsson$^1$ and J.~Dobaczewski$^{1,2}$ }

\affiliation{$^1$Department of Physics, Post Office Box 35 (YFL), FI-40014 University of Jyv\"askyl\"a, Finland}
\affiliation{$^2$Institute of Theoretical Physics, University of Warsaw, ul. Ho\.za 69, PL-00-681 Warsaw, Poland}

\date{\today}

\begin{abstract}
We extend density-matrix expansions in nuclei to higher orders in
derivatives of densities and test their convergence properties.
The expansions allow for converting the interaction energies
characteristic to finite- and short-range nuclear effective forces
into quasi-local density functionals. We also propose a new type of
expansion that has excellent convergence properties when benchmarked
against the binding energies obtained for the Gogny interaction.
\end{abstract}

\maketitle

%%%%%%%%%%%%%

The description of finite many-body systems is a continuous exciting
challenge in physics. In view of the exploding dimensionality of
many-particle Hilbert spaces, direct approaches are however often
untractable.
Instead, density functional theory (DFT) \cite{[Giu05]} provides an alternative by
showing that a much simpler description in terms of the density
alone is in principle possible.
 The main challenge
is finding or deriving the best
density functionals that are able to consistently describe large classes of
physical systems. In this work we address a related question, namely to what
extent the physics of the exact nonlocal exchange potential can be captured
by an approximate density functional?  This question is not only of
fundamental importance but also of great practical interest, and
attempts to solve it date back to the early 1950s,
involving the Slater as well as the optimized effective potential
approximations \cite{[Giu05]}.

For nuclear-physics applications, Negele and Vautherin developed
an alternative method known as the density-matrix expansion (DME) \cite{[Neg72]},
which
takes advantage of the short-range of the nuclear two-
and three-body forces.
It is particularly convenient
as it was originally constructed to
generate quasi-local
density functionals \cite{[Neg75a]} of a form similar to
the successful phenomenological Skyrme functionals
used extensively today \cite{[Ben03]}.
The
method have recently gone
through a revival \cite{[Fin06],[Bog09],[Geb09],[Dru10],[Dob10]}
and appears to be a promising starting point for the
construction of {\it ab initio} nuclear DFT models \cite{[Geb09],[Dru10]}.
The idea is
to start from effective forces, 
derived from first principles, and then transform the expression for the interaction energies
into quasi-local density functionals using the DME.
Because the starting point is an effective
force, this method is able to take correlations into account in a way
that goes much beyond the Hartree-Fock treatment of the bare
interaction.

 In conjunction with these {\it ab initio} methods, there
are empirical efforts to improve nuclear density functionals,
such as the systematic method we recently proposed to construct
nuclear density functionals using
higher-order derivatives of
densities \cite{[Car08e]}.
In such an approach, the DME is helpful as it can serve as a method of
classifying different terms according to their importance,
similarly to what is available in the
effective-field-theory power-counting scheme \cite{[Wei99]}.

In this Letter, we explore the ideas of the DME to transform %both
the direct and exchange energies resulting from the finite-range Gogny interaction
into quasi-local density functionals. We then extend these methods
beyond the second order expansions used so far to make the first tests of
convergence and accuracy in function of the expansion order.

%%%%%%%%%

We begin our analysis by recalling that the nuclear one-body density matrix
$\rho\left(\vec{r}_{1}\sigma_{1}\tau_{1},\vec{r}_{2}\sigma_{2}\tau_{2}\right)$,
which depends on the space-spin-isospin nucleon
coordinates $\vec{r}\sigma\tau$,
can be separated into four nonlocal spin-isospin densities
$\rho_{\vu}^{\vau}\!\left(\vec{r}_1,\vec{r}_2\right)$
as
\begin{eqnarray}
&&\!\!\!\!\!\!\rho\left(\vec{r}_{1}\sigma_{1}\tau_{1},\vec{r}_{2}\sigma_{2}\tau_{2}\right)=
\nonumber \\
&&~~~\tfrac{1}{4}\sum_{\vu\vau}\left(\sqrt{3}\right)^{\vu+\vau}
\left[\sigma_{\vu}^{\sigma_{1}\sigma_{2}},\left[ \tau^{\vau}_{\tau_{1}\tau_{2}},
\rho_{\vu}^{\vau}\!\left(\vec{r}_1,\vec{r}_2\right)\right]^{0}\right]_{0} ,
\label{eq:DensMat}
\end{eqnarray}
where we sum over the spin-rank $\vu=0$ and 1 (scalar and vector) and
isospin-rank $\vau=0$ and 1 (isoscalar and isovector) spherical tensors of the
Pauli matrices, $\sigma_\vu$ \cite{[Car10c]} and $\tau^\vau$, respectively. We
denote vector (isovector) coupling by the standard square brackets with
subscripts (superscripts) giving the values of the total
spin (isospin).

For a local interaction, the energy corresponding to the direct (Hartree) term depends on the
local densities $\rho_{\vu}^{\vau}\!\left(\vec{r}\right)\equiv\rho_{\vu}^{\vau}\!\left(\vec{r},\vec{r}\right)$
in Eq.~(\ref{eq:DensMat}) as
\begin{eqnarray}
\!\!\!\!{\cal E}^{\text{int}}_{\text{dir}}&\!\!\!=\!\!\!&\tfrac{1}{2}\!\!\int\!\!
\rmd\vec{r}_{1}\rmd\vec{r}_{2}\!\sum_{\vu\vau}\!V_{\vu\vau}^{\text{dir}}
\left(\vec{r}_{1},\vec{r}_{2}\right)
\left[ \left[\rho_{\vu}^{\vau}\!\!\left(\vec{r}_{1}\right),\rho_{\vu}^{\vau}\!\!\left(\vec{r}_{2}\right)\right]^{0}\right]_{0},
\label{eq:Hartree}
\end{eqnarray}
where $V_{\vu\vau}^{\text{dir}}\left(\vec{r}_{1},\vec{r}_{2}\right)$
denote the local direct two-body
potentials.
As discussed in Ref.~\cite{[Dob10]}, for short-range interactions and
sufficiently smooth
local densities one can employ simple Taylor expansions of densities
to obtain the energy density in the form of a gradient expansion.
As in Ref.~\cite{[Car08e]},
we use the spherical tensor representation of derivatives,
which allows for the most economical classification of terms in the
expansions.

Introducing the standard
average $\vec{R}=\frac{1}{2}\left(\vec{r}_1+\vec{r}_2\right)$ and
relative $\vec{r}=\left(\vec{r}_1-\vec{r}_2\right)$ coordinates, we have
the following expansion of the local densities,
\begin{eqnarray}
\rho_{\vu}^{\vau}\!\left(\vec{R}+\frac{\vec{r}}{2}\right)&\!=\!&
e^{\tfrac{1}{2}\vec{r}\cdot\hat{\vec{\nabla}}}
\!\!\rho_{\vu}^{\vau}\!\left(\vec{R}\right)
\!=\!\sum_{n}\tfrac{1}{n!}\left(\tfrac{1}{2}\vec{r}\cdot\hat{\vec{\nabla}}\right)^{n}
\rho_{\vu}^{\vau}\!\left(\vec{R}\right)\nonumber\\
&\!=\!&\sum_{nL}\,a_{nL}\,r^n\,\left[Y_{L}\left( \hat{r} \right),\hat{D}_{nL}\right]_{0}
\rho_{\vu}^{\vau}\!\left(\vec{R}\right),
\label{eq:DirectTay}
\end{eqnarray}
where, for clarity, the spin and isospin projection quantum numbers, $m_\vu$ and
$m_\vau$, are not shown, $\hat{\vec{\nabla}}$ denotes
the gradient operator with respect to variable $\vec{R}$, $\hat{D}_{nL}$
denote the spherical-tensor derivatives of order $n$ and rank
$L$ \cite{[Car08e]}, $Y_{L}\left( \hat{r} \right)$ are the standard
spherical harmonics, and $a_{nL}$ are simple numerical constants
resulting from the recoupling of the Cartesian $n$th order derivatives
$\left(\tfrac{1}{2}\vec{r}\cdot\hat{\vec{\nabla}}\right)^{n}$ into the
spherical-tensor forms \cite{[Car10b]}.
\begin{table}[ttt]
\begin{center}
\caption{\label{Tab:208Pb}
The direct end exchange energies in $^{208}$Pb calculated with the finite-range Gogny D1S
interaction \cite{[Ber91b]} in the scalar ($\vu=0$), vector ($\vu=1$),
isoscalar ($\vau=0$), and isovector ($\vau=1$) channels. All energies are in MeV.
}
\begin{tabular}{l@{\hspace{5ex}}r@{\hspace{5ex}}r@{\hspace{5ex}}r@{\hspace{5ex}}r}
\hline
\hline
& \multicolumn{2}{c}{$t=0$} & \multicolumn{2}{c}{$t=1$} \\
& $v=0$ & $v=1$ & $v=0$ & $v=1$ \\
\hline
${\cal E}^{\text{int}}_{\text{dir}}$ &  $-$12294.769 & 0      &   373.314 &  0     \\
${\cal E}^{\text{int}}_{\text{exc}}$ &    $-$594.992 & 0.629  & $-$32.337 &  0.215 \\
\hline
\hline
\end{tabular}
\end{center}
\end{table}

Expanded in this way, and after a partial integration and recoupling needed
to make all derivative operators act on the same density, the direct
interaction energy of Eq.~(\ref{eq:Hartree}) can
be written as a 
gradient expansion,
\begin{eqnarray}
\!\!\!\!{\cal E}^{\text{int}}_{\text{dir}}&\!\!\simeq\!\!&\!\!\sum_{mI\vu\vau}\int\!\!\rmd\vec{R}\, C_{mI\vu}^{\vau}
\left[ \left[\rho_{\vu}^{\vau}(\vec{R}),\left[\hat{D}_{mI},\rho_{\vu}^{\vau}(\vec{R})\right]_{\vu}\right]^{0}\right]_{0}.
\label{eq:Hartree2}
\end{eqnarray}
In this final form, the coupling constants $C_{mI\vu}^{\vau}$ are
numbers
obtained as moments of the
direct potentials
up to the order of the expansion $N$. This order is defined by keeping all
terms in the product that fulfill the condition
$n+n' \leq N$
, where
$n$ and $n'$ denote the orders in the expansions of the two densities
[compare with Eq.~(\ref{eq:DirectTay})].
\short{Lengthy explicit expressions for $C_{mI\vu}^{\vau}$ will be presented elsewhere
\cite{[Car10b]}.}

To test the quality and convergence properties of this expansion, we
generated a set of realistic densities for doubly
magic nuclei by solving the self-consistent equations
for the nuclear SLy4 functional \cite{[Cha98]}.
For these densities, the sums in Eq.~(\ref{eq:Hartree2}) were
evaluated to different orders by calculating the higher-order
derivatives of densities needed using the program
\pr{HOSPHE} (v\codeversion) \cite{[Car10c]}.
The coupling constants in Eq.~(\ref{eq:Hartree2}) were
calculated from moments of the finite-range part of the
Gogny D1S interaction \cite{[Ber91b]}. Then for the same
set of densities, exact energies were determined
using the capability of treating Gaussian interactions
with the code \pr{HFODD} (v2.45h)~\cite{[Dob09h]}.

\begin{figure}[ttb]
\includegraphics[clip,width=0.8\columnwidth]{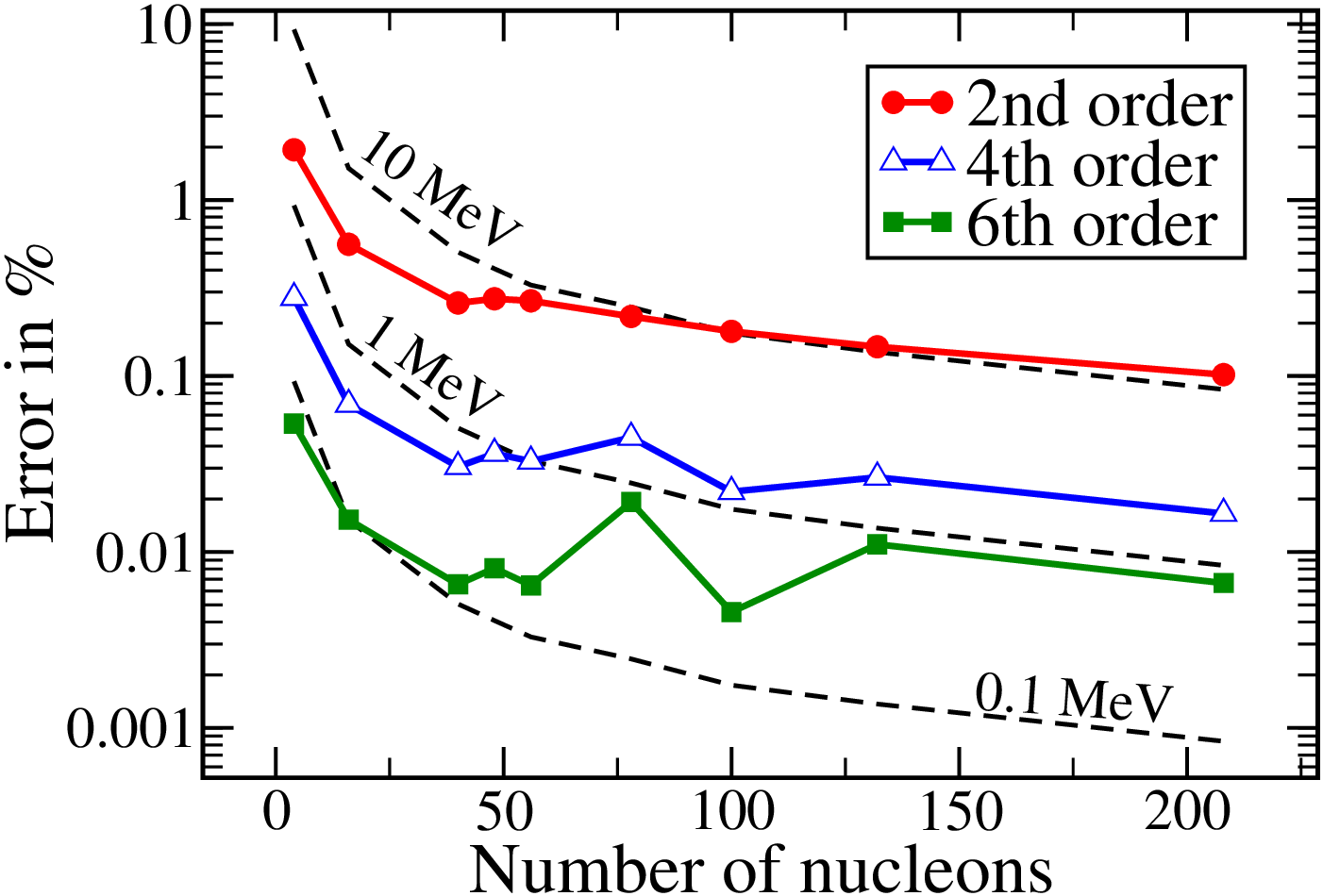}
\caption{\label{fig:DirectTerm}
(Color online) Precision of the 
gradient expansions
(\ref{eq:Hartree2}) of
the direct interaction energies calculated for the finite-range Gogny D1S
interaction \cite{[Ber91b]}. The nine nuclei used for the test are
$^4$He, $^{16}$O, $^{40,48}$Ca, $^{56,78}$Ni, $^{100,132}$Sn, and
$^{208}$Pb. Positive (fourth order) and negative (second and sixth
orders) errors are shown with open and closed symbols, respectively.
}
\end{figure}

As an example, in Table~\ref{Tab:208Pb} we show the exact energies
calculated in $^{208}$Pb for the four spin-isospin channels. In the
ground states of even-even nuclei,
the vector terms of the direct
energies vanish due to the conserved time-reversal symmetry, and thus
our calculations here only test the scalar terms of the expansion in
Eq.~(\ref{eq:Hartree2}). In Fig.~\ref{fig:DirectTerm} we show percentage
deviations between the gradient approximations and the exact results.
It is extremely gratifying to see that in light nuclei the sixth-order
expansion allows for reaching the precision of 0.1--0.01\% or 0.1\,MeV.
Even more important, in each higher order the precision increases by
a large factor, which is characteristic to a rapid power-law convergence.
One also sees that in heavier, isospin-unsaturated nuclei the precision
slightly deteriorates, probably owing to a slower convergence for
the isovector densities, which  fluctuate more than the isoscalar ones.
In comparison, models based on the best tuned Gogny forces give RMS deviations of
0.798 MeV \cite{[Gor09a]} for the description of experimental masses and to go
much beyond this precision in the expansion is thus not so useful.

One can, of course, quite easily calculate the direct terms exactly,
but still it is interesting to see that a few of the lowest-order coupling
constants 
appearing in the expansion
are able to capture the
essential physics contained in the nuclear finite-range force. For
the exchange (Fock) terms
whereto we turn now, the situation is
significantly different, because here the expansion into quasi-local densities
dramatically increases the simplicity and efficiency of calculations.

In order to deal with the exchange term, we propose
a new approximation, which we call a damped
Taylor (DT) expansion, and we define it in the following way:
\begin{widetext}
\begin{equation}
\rho_{\vu}^{\vau}\!\left(\vec{R},\vec{r}\right)\!=\!
\bar{\rho}_{\vu}^{\vau}\!\left(\vec{R},\vec{r}\right)+
e^{-\tfrac{r^{2}}{a^{2}}}
\left.\left[ e^{i\vec{r}\cdot\hat{\vec{k}}}
e^{\tfrac{\left(\vec{r}_{1}-\vec{r}_{2}\right)^{2}}{a^{2}}}
\!\!\left(\rho_{\vu}^{\vau}\!\left(\vec{r}_1,\vec{r}_2\right)-
\bar{\rho}_{\vu}^{\vau}\!\left(\vec{r}_1,\vec{r}_2\right)\right)
\right]\right|_{\vec{r}_{1}=\vec{r}_{2}=\vec{R}}
=\sum_{nLm}\pi_{nL}^m\left(r\right)\left[Y_{L}\left(\hat{r}\right),\rho_{nL\vu}^{\vau}\left( \vec{R}\right)\right]_{0}.
\label{eq:DT-DME}
\end{equation}
\end{widetext}
The Taylor expansion of the term in the square brackets is obtained by using
$
e^{i\vec{r}\cdot\hat{\vec{k}}}=\sum_{m}\frac{1}{m!}\left(i\vec{r}\cdot\hat{\vec{k}}\right)^{m},
$
where $\hat{\vec{k}}=\tfrac{1}{2i}\left(\vec{\nabla}_1-\vec{\nabla}_2\right)$
is the standard relative-momentum operator.
The quasi-local densities,
are defined as
\begin{equation}
\rho_{nL\vu}^{\vau}\left( \vec{R}\right)
=\left.\hat{K}_{nL} \rho_{\vu}^{\vau}\!\left(\vec{r}_1,\vec{r}_2\right)
\right|_{\vec{r}_{1}=\vec{r}_{2}=\vec{R}} ,
\label{eq:secondary}
\end{equation}
where the higher-order spherical-tensor derivative operators
$\hat{K}_{nL}$ \cite{[Car08e]} are built from the first-order
operators $\hat{\vec{k}}$.
Because of the simple form of the expansion, explicit formulas
for functions $\pi_{nL}^m\left(r\right)$ can be derived and
will be presented elsewhere \cite{[Car10b]}.

\begin{figure}[tbh]
\includegraphics[clip,width=0.8\columnwidth]{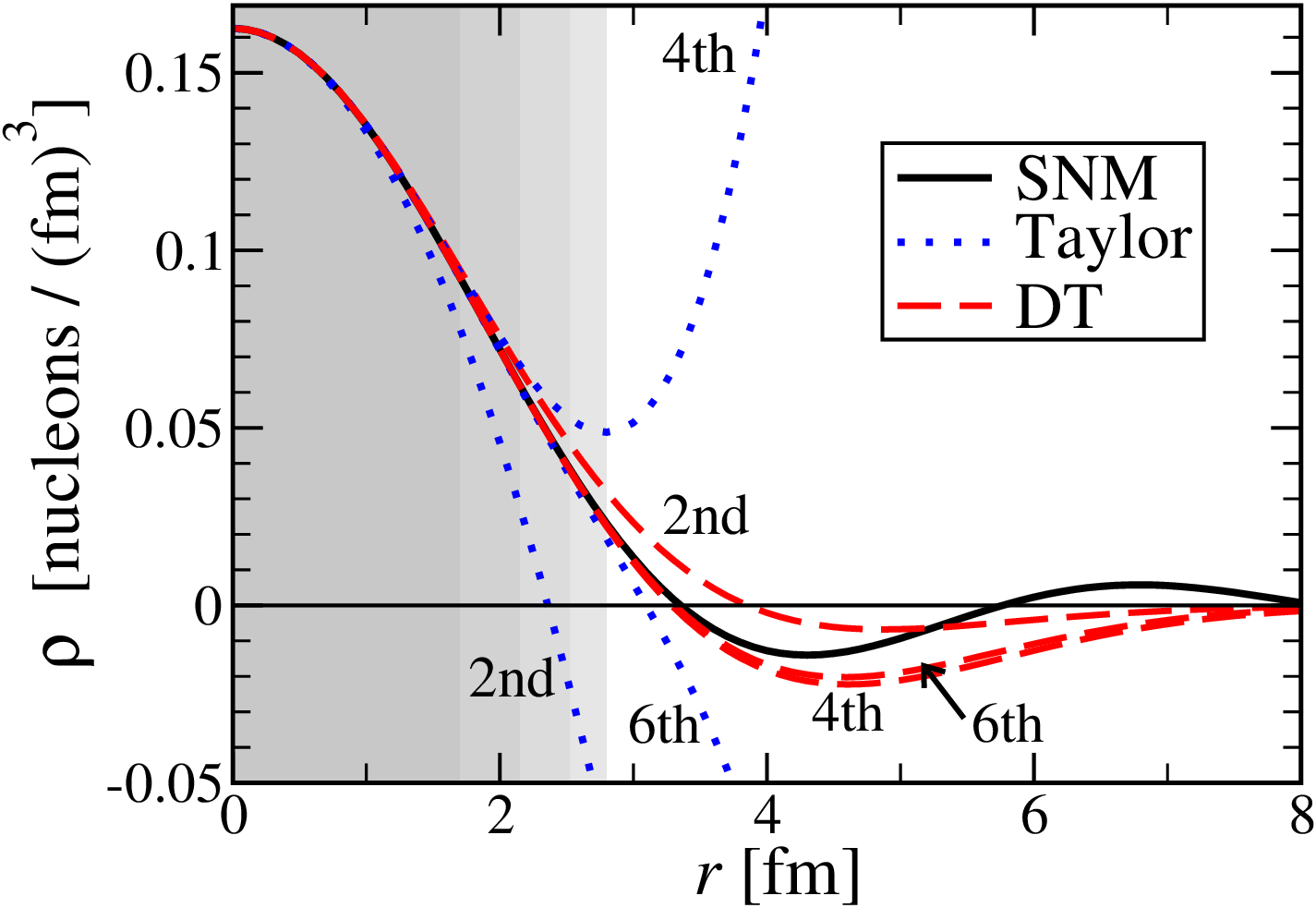}
\caption{\label{fig:NM}
(Color online) Comparison of a Taylor expansion and the DT expansion (with
$\bar{\rho}_{\vu}^{\vau}=0$ and $a=4/k_F$), applied to the SNM nonlocal density
in second, fourth, and sixth orders.
The figure is drawn for $k_F^0=1.35$ fm$^{-1}$ corresponding
to the equilibrium density of SNM \cite{[Ber91b],[Dob10]}.
The shaded regions
indicate the integration ranges needed to obtain 1, 0.1, 0.01 and
0.001 \% accuracy for the SNM exchange energy with the Gogny D1S
interaction.}
\end{figure}
The DT expansion has the potential to be more accurate
than the original Negele-Vautherin (NV) expansion \cite{[Neg72]} because
it avoids the approximation of angular averaging the density
matrix in the nonlocal $\vec{r}$ direction.
It is generated around
a model density $\bar{\rho}_{\vu}^{\vau}$ and is also
damped by a Gaussian factor with width $a$.
In the limit of vanishing model density
and infinite damping distance,
the DT becomes an ordinary Taylor expansion. This
Taylor expansion applied to the uniform system of symmetric nuclear matter
(SNM) is illustrated in
Fig.~\ref{fig:NM}.
It works well in a region around $\vec{r}=0$ but
diverges for larger relative distances.
These bad asymptotics can however be cured by
using a finite damping width $a$,
as also demonstrated in Fig.~\ref{fig:NM}. By
employing the
condition
that
it should work equally well in SNM for all values of
Fermi momentum $k_F$, we estimate this width as $a=4/k_F$.
Assuming that the densities of real
nuclei have similar behavior, we retain this estimate of $a$ in
the calculations presented below.

The model density
can be chosen in
different ways by the requirement that the truncated expansions
must become exact in certain limits. The model used in the following
leads to exact values in SNM, and is obtained by assuming that the density
in each point $\vec{R}$ has the SNM nonlocal dependence,
$\bar{\rho}_{\vu}^{\vau}\!\left(\vec{r}_1,\vec{r}_2\right)
=\rho_{\vu}^{\vau}\!\left(\vec{R}\right)\frac{3j_{1}\left(k_{F}r\right)}{k_{F}r}.$
For large relative distances, when the expansion around $r=0$ can no longer
be trusted, the bracketed term in Eq.~(\ref{eq:DT-DME}) is damped to
zero and the asymptotic behavior is then given entirely by this
model part. In this way the parameter $a$ interpolates
between placing trust in the model part or in the expansion part,
and the optimal balance is determined by the range in which the
expansion converges, as estimated above.

By inserting the DT expansion of Eq.~(\ref{eq:DT-DME}) into the expression for the
exchange energy, we obtain the quasi-local expansion
\begin{widetext}
\begin{equation}
\!\!{\cal E}^{\text{int}}_{\text{exc}}\!=\!\tfrac{1}{2}\!\!\int\!\!
\rmd\vec{r}_{1}\rmd\vec{r}_{2}\!\sum_{\vu\vau}\!V_{\vu\vau}^{\text{exc}}
\left(\vec{r}_{1},\vec{r}_{2}\right)
%\nonumber \\ &&\times
\left[ \left[\rho_{\vu}^{\vau}\!\left(\vec{r}_{1},\vec{r}_{2}\right),
\rho_{\vu}^{\vau}\!\left(\vec{r}_{2},\vec{r}_{1}\right)\right]^{0}\right]_{0}
\label{eq:Fock}
\simeq\!\!\!\!\sum_{\vau,nn'L\vu J}\!\int\!\rmd\vec{R}\, C_{nL\vu J}^{\vau,n'}
\left[ \left[\rho_{n'L\vu J}^{\vau}\!\left(\vec{R}\right),
\rho_{nL\vu J}^{\vau}\!\left(\vec{R}\right)\right]^{0}\right]_{0} ,
\end{equation}
\end{widetext}
where the
quasi-local densities $\rho_{nL\vu J}^{\vau}\!\left(\vec{R}\right)$, which
were introduced in Ref.~\cite{[Car08e]}, are equal to those of Eq.~(\ref{eq:secondary})
with the angular momenta $L$ and $v$ coupled to the total angular momentum $J$.
The coupling-constants
$C_{nL\vu J}^{\vau,n'}$ \short{, which in Eq.~(\ref{eq:Fock}) are
shown in the convention conforming to Ref.~\cite{[Car08e]},} are given
by the integrals
of the local exchange two-body potentials
$V_{\vu\vau}^{\text{exc}}\left(\vec{r}_{1},\vec{r}_{2}\right)
=V_{\vu\vau}^{\text{exc}}\left(r\right)$ and functions
$\pi_{nL}^m\left(r\right)$ of Eq.~(\ref{eq:DT-DME}).
They depend on
$k_F$ through the model
density
and damping width,
which in the standard local density approximation leads to a density dependence
$k_F=(3\rho_0^0/2\pi^2)^{1/3}$ \cite{[Neg72]}.
\begin{figure}[th]
\includegraphics[bb=1bp 30bp 772bp 523bp,clip,width=0.90\columnwidth]{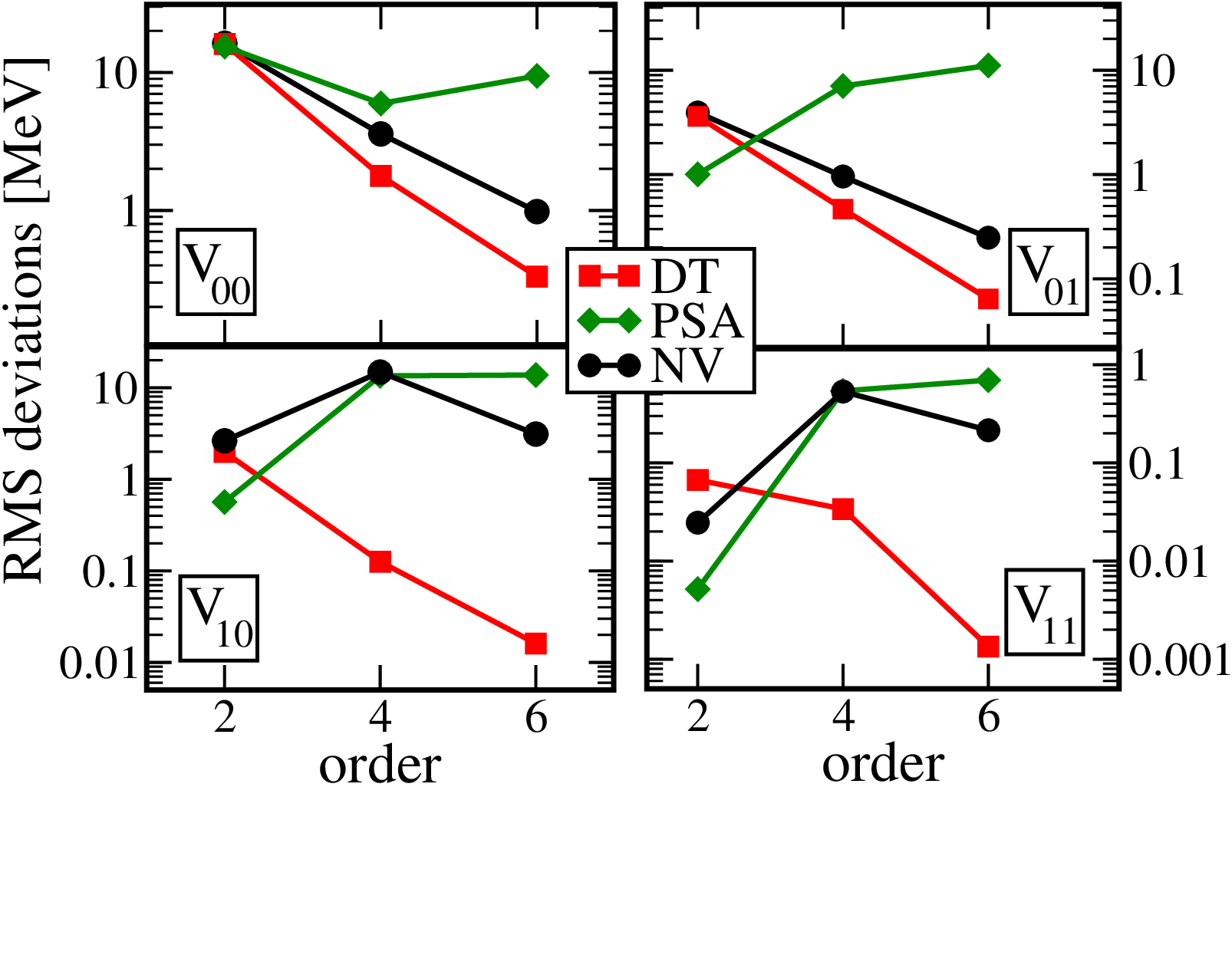}
\caption{\label{fig:Exchange}
(Color online) The RMS deviations between the exact and approximate exchange
energies of Eq.~(\protect\ref{eq:Fock}), calculated for the nine
nuclei listed in the caption of Fig.~\protect\ref{fig:DirectTerm}.
The four panels show
results obtained in the four
spin-isospin channels labeled by $V_{\vu\vau}$. Results obtained in
the present work (DT) are compared with those corresponding to the NV
\protect\cite{[Neg72]} and PSA \protect\cite{[Geb09]} expansions.
}
\end{figure}

In Fig.~\ref{fig:Exchange} we show results obtained with the DT
expansion
compared to the exact values, similar to
those listed in Table~\ref{Tab:208Pb}.
The comparison is made by considering the RMS deviations between the
exact and approximate exchange energies, and is also shown
for the NV \cite{[Neg72]} and
phase-space averaging (PSA) \cite{[Geb09]} expansions.
As seen in this figure,
the NV Bessel-function expansion
used in the scalar channels works very well, whereas the alternative
method used for the vector channels [see Eq.~(4.23) in Ref.~\cite{[Neg72]}]
does not improve when evaluated beyond the second order.

We employ the recently proposed PSA method in the INM-DME version \cite{[Geb09]} and
generalize it to higher orders by incorporating the model density in
the same way as in Eq.~(\ref{eq:DT-DME}). Other modifications are probably also
possible and could be studied as well. At second order, our
prescription reproduces that proposed in
Ref.~\cite{[Geb09]} with the PSA performed over SNM.
Fig.~\ref{fig:Exchange} shows that
this is the best expansion at
second order, where the total RMS deviation for the exchange energy is 14.09
MeV. When used for the vector part, the PSA expansion is similar to the NV expansion
and also has problems when evaluated beyond
second order.

\short{however, other extensions to higher orders are
probably possible and need to be studied.
The extension proposed here guarantees that the effect at any order is to reorder the
series such that $\pi_{00}^M$ always contain the proper
counter terms which ensures the exact values for SNM, as
discussed in Ref.~\cite{[Dob10]}.}

As seen in Fig.~\ref{fig:Exchange}, the DT expansion exhibits excellent
convergence properties both for the vector and scalar parts,
with every next order improving the results
by large factors. Note that in the vector channels, the exact
exchange energies of Table~\ref{Tab:208Pb} are quite small, so at
second order their relative description is poor. In contrast, at
higher orders their description becomes rather accurate.
The results obtained for
the DT expansion are not so sensitive to the value of the damping width
$a$, and values in the range of $a\approx3/k_F$ -- $5/k_F$ all
lead to improvements over the NV expansion.

The total RMS deviations obtained in the DT case for the
exchange energies at fourth and sixth orders are
1.54 and 0.36\,MeV, respectively.
Calculations using a fixed value of $k_F^0=1.35$\,fm$^{-1}$
give results that are not significantly different \cite{[Car10b]}.
They lead to
substantial simplifications, because then the density
dependencies of coupling constants can be neglected.

\short{At second order, the PSA expansion works well, and, as shown in
Ref.~\cite{[Geb09]}, in the vector channels it would have worked even better
had the quadrupole deformation of the Fermi surface been taken into
account. However, when extended to higher orders in the way we
proposed it here, it does not seem to converge. In the scalar
channels, the NV expansion works very well, whereas in the vector channels it
is similar to the PSA method and also has problems when extended to
higher orders.}

\short{Applying DME methods to the Gogny interaction allows us to look at the
correct scale of the interaction range which characterizes nuclear low-energy
phenomena. We demonstrate that DMEs can be constructed which converges
and give a good reproduction of binding energies.
The errors associated with these approximations have been
determined and quantified as functions of the order.}

In summary,
we presented the first analysis of density-matrix expansions extended beyond second order,
with derivatives of densities included up to sixth order. 
We also proposed a new type of expansion that has excellent
convergence properties.
The results demonstrate that its possible to
construct expansions into quasi-local density functionals
that converge for the description of low-energy nuclear
observables owing to the short range of nuclear forces.
These results also provide a baseline for phenomenological
adjustments of quasi-local higher-order nuclear functionals.

We
thank Johan Gr\"onqvist
for useful discussions.
This work is supported in part by the Academy of Finland and the University of
Jyv\"askyl\"a within the FIDIPRO programme,
and by the Polish Ministry of Science
under Contract No.\ N N202328234.

\end{document}